\documentclass[journal]{IEEEtran}
\usepackage[table]{xcolor}
\usepackage{bm}
\usepackage{mathrsfs}
\usepackage{amsmath,amssymb}
\usepackage{graphicx}
\makeatletter
\newcommand*\bigcdot{\mathpalette\bigcdot@{.8}}
\newcommand*\bigcdot@[2]{\mathbin{\vcenter{\hbox{\scalebox{#2}{$\m@th#1\bullet$}}}}}
\makeatother
\usepackage{pgfplots}
\usepackage{algorithm}
\usepackage{algpseudocode} 
\usepackage{caption}
\usepackage{subfigure}

\usepackage{epstopdf}
\usepackage{color}
\usepackage{cite}
\usepackage[T1]{fontenc}
\usepackage{multirow}
\usepackage{hhline}

\begin{document}

\title{Reconfigurable Intelligent Surface for Sensing, Communication, and Computation: Perspectives, Challenges, and Opportunities } 

\author{
       Bin~Li,
       Wancheng~Xie, and
       Zesong~Fei,~\IEEEmembership{Senior Member,~IEEE}


\thanks{Bin Li is with Nanjing University of Information Science and Technology, China (email: bin.li@nuist.edu.cn).}
\thanks{Wancheng Xie is with Xiamen University, China (email: zuoyeyiwancheng@gmail.com).}
\thanks{Zesong Fei is with Beijing Institute of Technology, China (email: feizesong@bit.edu.cn).}
}

\maketitle

\begin{abstract}
Forthcoming 6G networks have two predominant features of wide coverage and sufficient computation capability. To support the promising applications, Integrated Sensing, Communication, and Computation (ISCC) has been considered as a vital enabler by completing the computation of raw data to achieve accurate environmental sensing. To help the ISCC networks better support the comprehensive services of radar detection, data transmission and edge computing, Reconfigurable Intelligent Surface (RIS) can be employed to boost the transmission rate and the wireless coverage by smartly tuning the electromagnetic characteristics of the environment. In this article, we propose an RIS-assisted ISCC framework and exploit the RIS benefits for improving radar sensing, communication and computing functionalities via cross-layer design, while discussing the key challenges. Then, two generic application scenarios are presented, i.e., unmanned aerial vehicles and Internet of vehicles. Finally, numerical results demonstrate a superiority of RIS-assisted ISCC, followed by a range of future research directions.

\end{abstract}


\IEEEpeerreviewmaketitle

\section{Introduction}

The massive access of wireless devices and the requirements of higher data rate will demand a considerable amount of spectrum, which leads to serious spectrum congestion. To solve this challenge, the technique of Integrated Sensing And Communication (ISAC, a.k.a., dual-functional radar-communications)  which combines the communication and radar modules in the physical layer, has been suggested as an enabling technology via sharing the frequency band of radar and communication systems \cite{LiuFSPM2023,FLJSAC2022}, and further supports the sharing of hardware platform (such as antenna and radio frequency).

It is foreseen that the future 6G networks will not only serve 
the single purpose of \textit{reliable data transmission} from end to edge, but will need to support ubiquitous intelligence applications with the feature of \textit{high-accuracy of sensing} and \textit{low-latency of computation}, such as immersive extended reality, industrial Internet, and autonomous driving. For latency-sensitive sensing tasks (e.g., disaster rescue), the long computation latency of sensing data may dramatically reduce the value of sensing results and further lead to serious safety incidents. Therefore, 
there is a surge of interest to explore the converging functionalities of sensing, communication, and computation, which is referred to as Integrated Sensing, Communication, and Computation (ISCC) \cite{FengZNetwork2021}.
The rationale is that radar sensing, communication, and computation functionalities are tightly converged and can be simultaneously realized by using the same hardware module. In fact, the adoption of ISAC has
prompted the development of ISCC since it allows efficient sensing data acquisition and task offloading with the help of Mobile Edge Computing (MEC).
In this context, multiple ISAC devices first perform radar sensing to obtain multi-view data, and then upload the sensory data to edge servers to enable low-latency intelligent services \cite{DingJSAC2022}. 
Specifically, the target detection is accomplished by processing echo signal at radar receiving antenna array, the data transmission is completed by processing the received signal at the receiving antenna array, and the uploaded input data is timely computed by edge servers to make
efficient decision.

The study of ISCC is currently in nascent stage, and there have been very limited related works. For instance, \cite{LZTWC2022} proposed a wireless scheduling architecture by considering the inherent conflicts among communication, sensing and computation performance. \cite{QQTWC2021} developed an ISCC architecture where the non-orthogonal multiple access channels are used for sensing.  The work in \cite{DingJSAC2022} designed a multi-objective optimization problem for ISCC by jointly considering computation energy consumption and radar beampattern.
However, the severe wireless propagation path loss greatly restricts radar sensing and data processing capabilities. 
The echo signal received by the radar-aided access points is weak due to significant channel attenuation, which degrades the performance of the ISCC system. 

\begin{figure*}[!t]
	\centering
	\includegraphics[width=18cm]{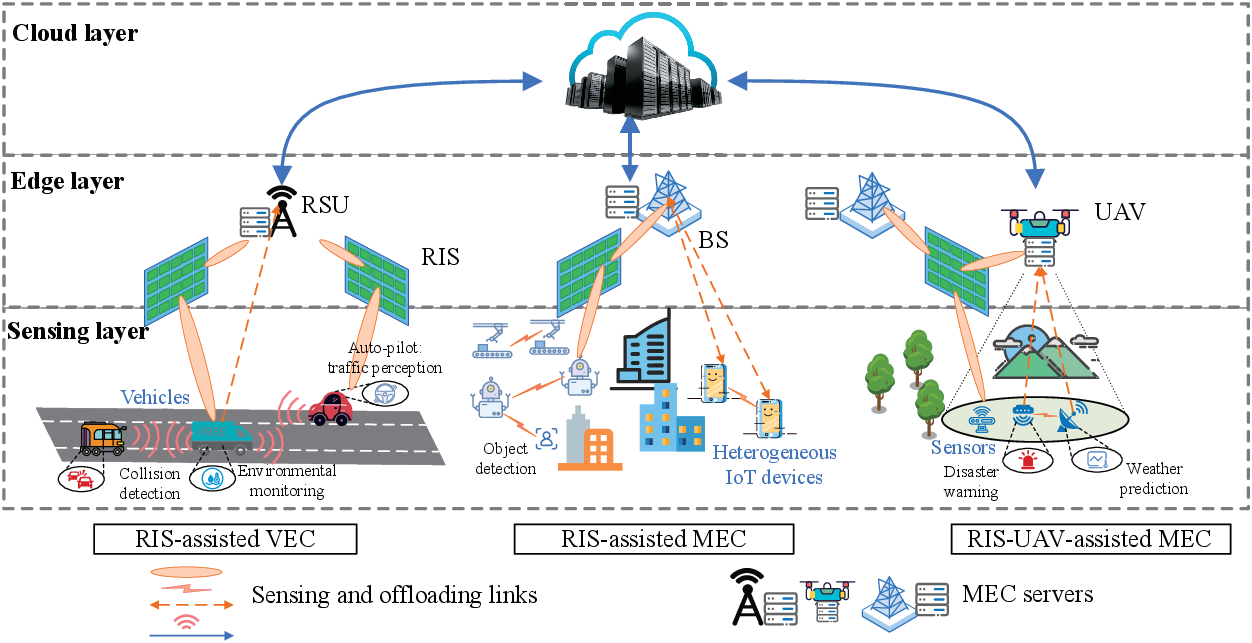}
	\caption{An illustrative scenario of the considered ISCC network, where some IoT devices may reach the edge nodes (i.e., BS, RSU, and UAV) through an indirect path supported by RIS with the configurable phase shifts.}
	\label{sys}
\end{figure*}

Recently, Reconfigurable Intelligent Surface (RIS) has drawn growing interest due to its ability of favorably adjusting the wireless propagation environment. With the assistance of RIS, the weak or blocked direct channel for sensing or computation offloading can be remarkably enhanced, as the channel along the path from source to RIS	and then to the destination is modeled as the two-phase cascading channel \cite{BaiWCM2021}. Unlike Multiple-Input Multiple-Output (MIMO) relay and massive MIMO, RIS comprises of dozens of low-cost passive reflect elements and does not harness any radio frequency chains, which can reduce hardware and energy footprints. In addition, it can be readily installed on the walls, the ceiling of a large indoor area, and the facades of a building, which is recognized as a key technology to breakthrough the restriction of traditional wireless communications. Although a few works have been carried out on enhancing the performance of ISAC systems, there is no prior work using RIS to facilitate the performance improvement of ISCC networks. Additionally, the classic alternating optimization algorithm for finding a suboptimal solution cannot be directly adapted to dynamic network conditions.

To provide an in-depth understanding, in this article,
we propose an RIS-aided ISCC framework, in which RISs are leveraged for coverage extension of computing service and sensing capability enhancement via additional spatial freedom. The Deep Reinforcement Learning (DRL) algorithm is exploited to design an RIS-aided ISCC scheme.
\begin{itemize}
	\item We first present a systematic overview of RIS-enabled ISCC to address the spectrum scarcity problem and suppress the serious channel fading. Some key challenges for realizing the ISCC framework are discussed.
	
	\item We then illustrate two potential applications for ISCC including Unmanned Aerial Vehicle (UAV) and Internet of Vehicles (IoV), and elaborate the benefits of ISCC for the two networks.
	
	\item  Subsequently, we present a use case scenario enabled by RIS and give the evaluation results of synergizing RIS and ISCC through DRL algorithm. Numerical results show the proposed RIS scheme can reduce the energy consumption compared to the conventional scheme.
	Finally, we highlight the future research directions.
\end{itemize}

\section{RIS-Enabled Multi-Functional Networks}

The development of the current network technology has not been divorced from the basic paradigm of a single system, which has greatly restricted the innovation demand of the network technology and service provisioning.
With the continuous integration of information and communication technologies, we can envision that the 6G mobile network will 
become a multi-functional network that integrates sensing, communication, and computation functions.
The performance of ISCC is passively constrained by 
the non-favorable propagation conditions because of the dynamic environment and blocking events.
It is meaningful to deploy RIS in ISCC system to pursue performance improvement with new spatial degrees of freedom by controlling signal propagation, opening up new opportunities. 
\begin{itemize}
	\item RIS is capable of extending the coverage and connectivity of wireless networks by adaptively shaping propagation environments, and it is thus helpful for uplink
	and downlink communications.
	
	\item RIS can be applied in high-precision sensing systems, where the signal quality between radar transceiver and sensing target can be strengthened by smartly manipulating the phase shift of reflecting elements.
	
	\item RIS also enhances the uplink offloading success rate, where both the latency and the energy consumption of MEC systems are saved.
\end{itemize}

\begin{figure*}[t]
	\subfigure[RIS passive sensing]{  
		\begin{minipage}{6cm}
			\centering    
			\includegraphics[scale=0.5]{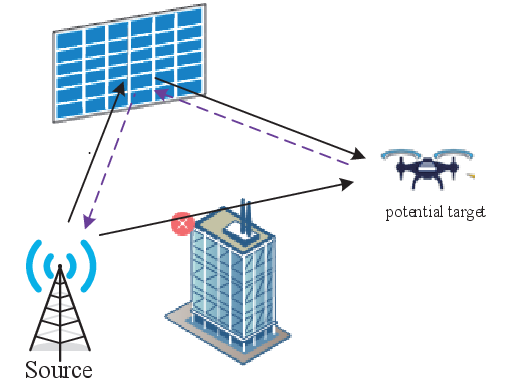}  
		\end{minipage}
	}
	\subfigure[RIS semi-passive sensing]{ 
		\begin{minipage}{6cm}
			\centering    
			\includegraphics[scale=0.5]{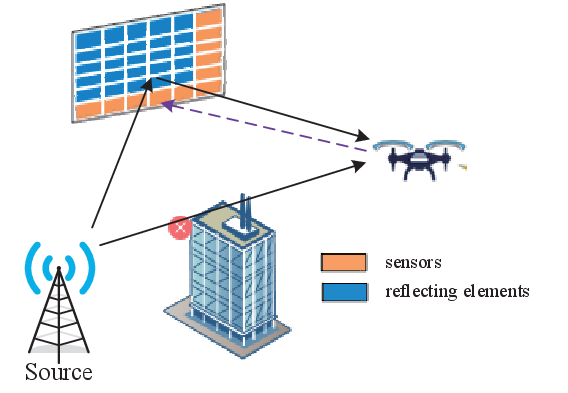}
		\end{minipage}
	}
	\subfigure[RIS active sensing]{ 
		\begin{minipage}{6cm}
			\centering   
			\includegraphics[scale=0.5]{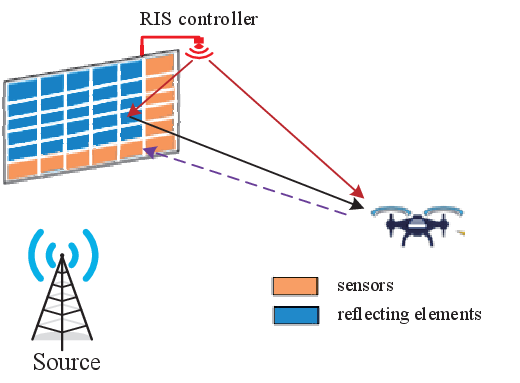}
		\end{minipage}
	}
	\caption{Typical RIS-aided sensing situations.}  
	\label{fig:three}    
\end{figure*}

Although the integration of communication and sensing, and the integration of computing and communication have been studied in recent years, ISCC is still in its infant stage. This section illustrates an RIS-enabled ISCC multi-functional network architecture as shown in Fig. \ref{sys}.
With the RIS, the channels from IoT devices to the edge nodes (i.e., BS, RSU, UAV) contains both the direct link (IoT device-edge node link) and the reflected link (IoT device-RIS-edge node link), where the direct link is characterized by both Line-of-Sight (LoS) and non-LoS, and the reflected link includes the IoT device-RIS link, the phase shifts at RIS, and the RIS-edge node link.

\subsection{RIS for Sensing and Communication}

As the two main applications of modern electromagnetic theory, radar and communication develop independently and gradually upgrade to multi-function equipments. The future paradigm will entail not only high-rate communication, but also high-resolution sensing services. Based on the strong similarity between radar system and communication system, extensive research on ISAC has been conducted to diminish the congestion of spectrum resources by using the unified waveform \cite{Liyanaarachchi2021}. In ISAC systems, the sensing and communication applications can benefit from sharing information with each other.

\textit{Sensing-enabled communication:}  
The sensing function provides prior information for communication by obtaining more plentiful user information and environmental data.

\textit{Communication-enabled sensing:} The communication function can effectively deliver and aggregate sensing information to support multi-node collaborative sensing, and then expand the dimension and depth of sensing.

Besides improving communication performance,
RIS can be deployed as a flexible ``bridge'', i.e.,
an additional path, between radar and targets to achieve larger sensing coverage and enhance sensing performance.
By applying RIS in ISAC systems, the transmit probing signal will
reach the target via the direct link and the RIS reflected link.
From the radar detection perspective, a widely adopted criterion is guaranteeing the radar INR requirement, which can be promoted by jointly designing RIS reflection coefficient and transmit beamforming \cite{XWang2022}. To this end,
through proper modeling and optimization design of RIS reflection coefficients, the wireless transmission environment of the system can be intelligently reshaped, and the ISAC performance can be better coordinated and improved.
Fig. \ref{fig:three} illustrates a few typical RIS-aided sensing situations including RIS paasive sensing, RIS semi-passive sensing and RIS active sensing, which are categorized based on the RIS's role \cite{ShaoXJSAC2022}.

\subsection{RIS for Communication and Computation}

Massive IoT devices are envisioned to be interconnected, while inspiring a vast amount of data from various services and applications (e.g., virtual reality and smart manufacturing). However, due to the limited computing resource and battery supply, smart devices are typically incapable of smoothly accommodating the resource-intensive services and applications if only relying on their local computing.
To address the computation burden, MEC empowered with sufficient computation resource at the network edge, has been recognized as a prevailing solution to save energy and guarantee low-latency services.
As such,  the integration of computing and communication has gradually deepened. 

Wireless transmission is indispensable for computaion-intensive devices to realize task offloading, however, task offloading may encounter the performance bottleneck that directly affects the uploading/downloading delay. On another front, the direct connections between the users and the edge servers may be frequently blocked by the obstacles. 
As a seamless integration of RIS and MEC, RIS-enabled MEC is regarded as a prominent solution to assist the communication and provide enhanced computation performance for delay-sensitive and data-intensive applications.

In order to improve the resource utilization and the computing performance, the computation resource and communication resource as well as RIS phase shift should be carefully designed. 
The work in \cite{BaiTJASC2020} attempted to introduce RIS into MEC networks
to minimize the latency by jointly optimizing the volume of
task offloading, the computing resources of edge servers, and the
phase shifts of RIS.
In \cite{CaoNetwork2021}, an RIS-assisted collaborative MEC architecture was proposed in space-air-ground networks to improve system capacity and reduce network latency.

\subsection{RIS for Sensing, Communication, and Computation}

The ISCC is an effective framework to realize the intelligent interconnection of human-machine-object and the efficient intercommunication of agents, which provides an important support for the innovation and development of 6G networks. The main feature is that it can deeply integrate the sensing technology with the wireless communication technology, and carry out auxiliary computing processing with the help of widely distributed computing capabilities. 
In this way, the wireless networks can realize high-precision, fine-perception and low-latency with large bandwidth. Compared with dedicated sensing and communication functionalities, ISCC offers two main advantages: 

\emph{Integration gain:} The gain in integration can significantly strengthen the spectrum utilization efficiency and reduce the hardware costs via sharing resources effectively.

\emph{Coordination gain:} The gain in coordination is to balance the radar sensing performance and the uplink offloading transmission performance in MEC. 

\begin{figure}[!t]
	\centering
	\includegraphics[width=9cm]{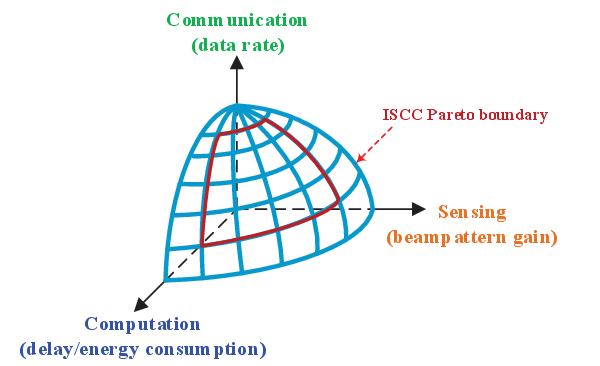}
	\vspace{-4pt}
	\caption{The performance region of ISCC.}
	\label{3D}
\end{figure}

In the proposed RIS-enabled ISCC architecture, IoT devices perform radar sensing and communication using the same hardware and signaling, while MEC is
introduced to assist the computation for radar sensing data.
Specifically, IoT devices first sense the environmental information by analyzing the raw reflected radio signals from the sensing targets (e.g., health monitoring and human motion recognition),
and achieve the functions of detection, ranging, positioning, and so on. The wealth of sensing data is usually computation-intensive and needs to be offloaded through wireless links and processed by taking full use of the computing power at edge servers. A major hurdle to overcome the double propagation loss during radar sensing and
task offloading is the high power consumption.
An RIS-assisted communication scheme to support the task offloading by exploiting the link-enhancing potential of RIS is preferred. Specifically, RISs are placed in a region to enhance the wireless communication links, and each RIS comprises $M$ passive reflecting elements which could manipulate the phase of the incident signal waves.
By jointly considering the task offloading decisions, transmit power of IoT devices, and phase shift of the RIS, the user quality-of-service (QoS) and user-perceived quality-of-experience in terms of sensing accuracy and computation latency can be enhanced.

The performance region of ISCC has a higher dimension
than those of the isolated multi-user communication, radar sensing,
and task computation.
The achievable performance region for ISCC is shown in Fig. \ref{3D}.
Particularly, 1) sensing
and communication compete for radio resources, and the communication
technique (e.g., 5G and beyond) further determines the required computation level since the quantized features can be transmitted reliably to the edge cloud for computing. 
2) the real-time accurate sensing can be enabled by combining distributed MEC and efficient information delivery. 3) the enhanced sensing function can provide prior information for the fast scheduling of distributed MEC and also provide more abundant data sources for intelligent services to enhance the robustness of training model, while the enhanced communication function can further improve the ubiquitous computing capability of the MEC network.
To this end, the three processes are highly coupled and thus need to be jointly considered to fully unleash its potentials. 
This thus calls for a joint analysis from a systematic view of integrating sensing, computation, and communication. The current performance metrics of interest for the ISCC system is data rate for communication, sensing accuracy (high-quality localization) and latency/energy consumption for task offloading. To facilitate understand the interaction among sensing, communication, and computation, we formulate three typical multi-objective optimization to specify the trade-offs among them. By jointly optimizing the user association, transmit beamforming, task offloading and computation resource allocation, the primary objective is to \\
 1) maximize the data rate while guaranteeing the sensing latency and quality requirements; \\
 2) maximize the sensing beampattern gain while satisfying the communication QoS and minimum delay requirement; \\
 3) minimize the latency/energy consumption while guaranteeing the QoS requirements on sensing, computation, and communication, respectively.

\begin{table*}[!t]
	\centering
	\setlength{\tabcolsep}{2pt}
	\renewcommand{\arraystretch}{1.8}
	\caption{Qualitative Comparison of ISAC, MEC and ISCC Networks.}
	\begin{tabular}{| c | p{3.5cm} |p{3.8cm} | p{3.8cm} |}
		\hline
		\textbf{Networks} & \textbf{Features} & \textbf{Benefits} & \textbf{Challenges}   \\
		\hline
		\hline
		ISAC    &  sensing and communication simultaneously  & improve communication and sensing dual fucntions  & the hardware and signal processing capabilities on the terminals and network capacity
		  \\
		\hline
		MEC       & communication and computing simultaneously& improve data rate and reduce computation latency  & the compuation-limited resource and unstable wireless channels    \\
		\hline
		ISCC   & sensing, communication and computing simultaneously & improve detection accuracy and reduce computation latency  & performance tradeoff among the data rate, sensing accuracy and computing efficiency      \\
		\hline
	\end{tabular}
	\label{comparison}
\end{table*}

As a summary of this section, we list the comparison of different RIS-enabled networks in Table I, including RIS-enabled ISAC, RIS-enabled MEC, and RIS-enabled ISCC.

\subsection{Key Challenges}
The joint application of RIS and ISCC techniques has the potential to address the fundamental performance limitations of intelligent IoT. Though promising, the implementation of RIS in ISCC faces new challenges which are described below:

\emph{1) Sharing Waveform Design:}
The transmitting waveform sharing demands to complete the functions of data transmission and radar detection at the same time, which not only needs to consider the theoretical performance of radar and communication, but also needs to pay attention to the complexity of hardware implementation, power efficiency and other implementation problems. Many waveform sharing approaches have been proposed in the current ISAC research, but it is difficult to determine which waveform sharing approach is more suitable for which scenario due to the lack of unified measurement standards. 
As a newly-increasing research field, the waveform design faces great challenges in terms of computational complexity and hardware cost.

\emph{2) Service Continuity:}
Service continuity refers to the functional continuity of sensing, communication, and computation. Sensing continuity refers to the ability of continuous information collection, localization, and tracking targets. Communication continuity refers to the handover of access points. Computation continuity refers to the ability to decompose and migrate computation tasks in real-time due to the dynamic changes of computing resources. However, the prior information required for optimization is not easy to obtain in highly dynamic change environments (e.g., IoV).

\emph{3) Hardware Limitations:}
The sensing accuracy is mainly dominated by the high precision locating of the
devices. Higher sensing accuracy may need to take into account the additional signal information such as angle of arrival, direction of arrival, multipath components, etc. 
For accurate sensing and communication, the estimation of channel parameters is impacted by the hardware impairments and the estimation process needs to be able to separate the physical channel from hardware.
The research challenge is to understand how hardware impairments impact the sensing accuracy, and how the signals should be designed to make estimation robust against hardware impairments.

\emph{4) Performance Tradeoff:}
For ISCC, wireless resources are used on a shared platform, this naturally leads to a highly coupled process among sensing, communication, as well as edge computing. For instance, the congested wireless channels will deteriorate the delivery performance of radar sensing results due to severe mutual interference, and the overloaded edge servers can lead to long computing latency and poor inference accuracy. Due to the inherent conflicts between targets, the optimization of single target is at the cost of the deterioration of other targets, in a way that it is difficult to have a unique optimal solution. Instead, a multi-objective optimization or tradeoff is incured among them to make the overall target as optimal as possible.

\section{Applications of Multi-Functional Networks} 

The ISCC networks can be deployed on access points, BSs, road side units, or UAV platforms to realize sensing, communication and computation functionalities simultaneously.
UAV and IoV networks as two typical application scenarios will be outlined in this section. 

\subsection{ISCC in UAV Networks}

Compared with the terrestrial networks, UAV-assisted
wireless networks have the potential to cover a larger area
with a higher data rate, owing to their high mobility and flexible deployment as well as the strong LoS propagation channels for the air-to-ground links, which is a natural fit for communication and sensing \cite{AzariMag2022}, such as  precision agriculture and disaster management.
On one hand, UAV that is equipped with built-in radar sensing devices can fly to remote area to carry out sensing missions. On the other hand, UAV can be deployed in the role of relay over the complex terrains and provides reliable communication links for the ground IoT devices, while the UAV receives the echos of the sharing waveform and performs radar detection to sense the target area. 

With the wide penetration of UAV in multifarious fields, high-performance UAV has been viewed as edge servers in the sky with sensing, computing and storage capabilities, which can be deployed immediately on-demand and provides
ISCC services to emergence scenarios and temporary hotspot situations \cite{WYJSTSP2021}. If the sensing, communication and computing modules as well as algorithms are integrated on UAV, the UAV-based ISCC architecture is explored for the development and implementation of next generation technologies.

In emergency communications, the rescue gradually develops from single UAV detection to multi-UAV cooperation.  The cooperative detection between UAVs requires that UAVs can not only perceive the surrounding environment and target, but also complete information interaction with cooperative UAVs. 
In order to accommodate the multi-functional requirements of complex battlefield, UAVs need to be equipped with communication, sensing and computing platform, which will increase the load weight and the electromagnetic interference, diminish the maneuvering performance of UAV. In ISCC-enabled UAV systems, multiple UAVs conduct data transmission and target detection at the same time, the mutual interference of radar echo and communication signal may occur. In other words, at the receiving ends of radar, it is necessary to suppress the communication signal and accurately obtain the target parameter information. At the receiving ends of communication, it is necessary to suppress the target echo and accurately recover the original information carried in the signal with interference. As such, the waveform design based on orthogonal multiplexing signal becomes the mainstream of mutual interference cancellation.

\subsection{ISCC in IoV}
IoV has been recognized as one major vision of the future intelligent transportation. 
In IoV, vehicle-to-vehicle (V2V) and vehicle-to-infrastructure
(V2I) communications can help acquire real-time traffic state to reduce traffic congestion and improve traffic safety,
where each vehicle could monitor its neighboring traffic conditions by vehicle-mounted radar. Then, each vehicle can reliably deliver the sensory data via V2V and V2I communications to the resource-rich edge servers where the data is timely processed for estimating road conditions. 
Future IoV systems feature a larger number of devices and multi-access environments where emerging services demand unprecedented high accuracy, ultra-low latency, and large bandwidth \cite{CLJSAC2022}. However, these services have a high variance in their resource demand with respect to time, location, context, as well as individual patterns. In order to realize a more intelligent transportation system, efficient sensing, communication and computation functionalities are tightly demanded simultaneously.

ISCC provides a new design idea for the development of IoV, especially in the intelligent transportation applications 
which requires the simultaneous realization of vehicular environmental sensing and computation offloading. 
The proper system platform enables all vehicles on the road to interact in a cooperative radar sensor network, providing unique safety features and intelligent traffic routes. Also, due to the radar sensing function, the ISCC system is not affected by the sun, rain, snow, fog and other complex weather, and can measure and sense the traffic environment all-time and all-weather. 
Furthermore, the emerging computation-intensive applications,
such as real-time driving monitoring, video entertainment, and
vehicle-road coordination,
leading to high communication and computing latency and unguaranteed QoS. A cloud-edge network architecture should be adopted to upload data to the edge servers or the cloud service platform for effectively solving the offloading delay problem.

The existing radar and communication composite waveform design methods have not taken into account the characteristics of vehicle-mounted platform and IoV environment, thus are incapable of achieving high transmission rate and good anti-multipath fading performance simultanously for communication functions. In the future intelligent transportation system, the traffic environment sensing ability of the vehicular networks needs to be further improved without affecting the communication performance, while the performance of radar target parameter estimation is not affected by the design of communication signal parameters. Additionally, the rapid channel change caused by the high mobility of vehicles makes the uncertainty in task offloading. Thus, the sensing and MEC strategies suitable for IoV should be jointly designed for providing safe autonomous driving.

\section{Case Study: Exploiting RIS for ISCC}

In this section, we first give a case study scenario and 
then evaluate the proposed RIS-assisted ISCC architecture.

\subsection{Scenario Setting} 
We consider a scenario consisting of a BS with 16 antennas, an RIS with 40 reflect elements, and 16 ground users. Each user embeds a radar with 16 antennas for sensing process. The location of BS, the RIS, and the center of ground users are set to be [-200, 0] m, [0, 0] m and [0, 30] m, respectively. The size of computational data is set between 0.4 Mb and 0.5 Mb, the average CPU cycles per bit is set between 800 and 1000, and the computational resource of users and BS is set as 1 GHz and 10 GHz, respectively. The bandwidth is set to 2 MHz, the noise power is set to -115  dBm, the transmission power budget is 0.5 W, and the path loss exponents of direct and reflect links are 3.6 and 2.2, respectively.
Aiming at the energy consumption minimization, we utilize DRL method to optimize the settings of RIS reflect elements, radar beampattern design, and offloading decisions under the latency constraints. The essential components of Markov decision process are introduced as follows, which can guide the agent in making optimal decisions.
\begin{enumerate}
	\item \textit{State}: The DRL agent observes the state of environment, and inputs the state into its policy network. The state space includes the size of computational task, the channel information and the location of users. 
	\item \textit{Action}: After obtaining the state, the agent gets its action by policy network. The action space includes the offloading decision, the RIS phase-shift and the beamforming vectors.
	\item \textit{Reward}: The agent executes the action and gets reward from environment. The reward is defined as the energy consumption of users.
\end{enumerate}

\vspace{-10.pt}
\subsection{Results Evaluation} 

\begin{figure}[!t]
	\centering
	\includegraphics[width = 9cm,height= 7cm]{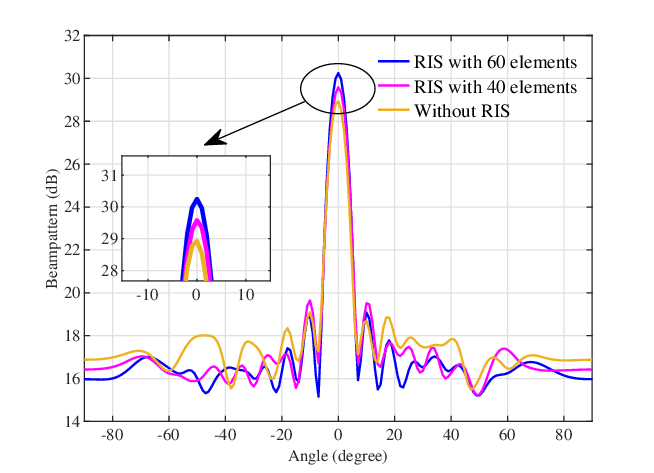}
	\caption{Radar beampatterns obtained with and without the aid of RIS.}
	\label{fig3}
\end{figure}

\begin{figure}[!t]
	\centering
	\includegraphics[width = 9cm,height= 7cm]{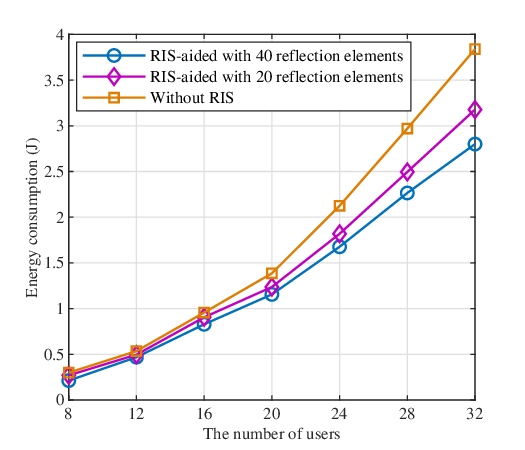}
	\caption{Energy consumption versus the number of users.}
	\label{fig4}
\end{figure}

This subsection presents numerical results to show the performance that can be achieved by RIS.
We first present the comparison of the radar transmit beampattern between with RIS and without RIS under different angles. The desired direction of radar is set as 0. It can be seen from Fig. \ref{fig3} that the desired direction has the maximum beampattern. Moreover, the schemes with RIS-aided have larger beampattern on desired direction than that of the scheme in which no RIS is exploited. This is because RIS is able to help decrease the interface between users. Hence, it is demonstrated that RIS is able to improve the sensing performance. 

We then evaluate the total energy consumption under different number of users in Fig. \ref{fig4}. Intuitively, it can be observed that the energy consumption of users gradually increases when the number of users grows. Moreover, the RIS-aided schemes obtain lower energy consumption, and as the RIS reflection elements increase, the performance becomes better. That is because when more reflection elements are equipped on RIS, the offloading links become stronger and the transmission rate gets faster, thereby leading to a lower transmission latency and energy consumption. Therefore, it can be  concluded that the ISCC scheme can effectively enhance system performance.

\section{Future Research Opportunities}
In this section, we shed light to several potential research opportunities aiming to compose a constructive integration of emerging technologies into ISCC networks.

\subsection{Distributed and Collaborative ISCC} 
With the benefits of combining RIS and UAV, by deploying multiple RISs and UAVs in the hot spot or edge areas, the wireless devices that are capable of ISCC can be ensured high information transmission rate and high-precision sensing, as well as enhanced coverage of computing services.
The performance of far-field RISs-aided ISCC networks is currently in nascent stage, the near-field RISs have also gained much research attentions, the near-field RISs need to be further investigated to exploit the potentials of ISCC networks.

\subsection{Digital Twin for ISCC} 

More powerful AI technologies can be supported by digital twin to provide users with timely decisions in the optimization of service quality. In ISCC, digital twin can
replace the users and edge servers to make offloading decisions in the virtual space in advance,
while the computing and communication resources between users and edge servers in the physical
space can be provided quickly and accurately. In
fact, digital twin serves as a potential solution to help deal with the highly dynamic and unpredictable network nature, thus strengthening computation offloading and communication decisions, which is of paramount significance to the development of ISCC networks.


\subsection{ISCC Networks at High Frequency Bands} 

Due to the large accessible bandwidth, Terahertz (THz) frequencies have the promise of high-precision sensing and extremely low latency, especially in virtual reality services. ISCC networks at THz frequencies are expected to reduce the complexity of beam training and improve the communication performance through the assistance of sensing information.
However, the wireless signals at THz frequencies will suffer from severe propagation attenuations, which is more suitable for short-range ISCC scenarios.
As such, efficient approaches (e.g., RIS and massive MIMO) should be considered to enhance the propagation distance via creating virtual LoS paths,  thereby reducing operating costs and 
improving the quality of ISCC networks at high frequencies.

\subsection{Sustainable ISCC Networks} 

In order to complete the transmission and computation of massive data, both the access sides and the sensing devices have to consume lots of energy, which seriously affects the sustainable execution of tasks in ISCC networks.
The green ISCC network aims to save energy and reduce the carbon emission from three point of views, including green communication technology, green sensing technology, and green computing technology, which call for further study.

\section{Conclusion}
In this article, we propose an RIS-enabled ISCC architecture where RIS combining MEC is introduced into a dual-function radar communication system. Then, the key implementation issues are discussed. Especially, two application scenarios are presented, namely UAV and IoV. Furthermore, a case study and performance evalution are provided, and the superiority of RIS-aided ISCC is verified compared with the no-RIS scheme. This article sheds the light onto the new research area of ISCC and outlines the essential research directions.

\section*{Acknowledgment}
This work was supported in part by the National Key R\&D Program of China (No. 2021YFB2900200), in part by the National Natural Science Foundation of China (No. 62101277), and in part by the Key Laboratory of Cognitive Radio and Information Processing, Ministry of Education (No. CRKL230203).

\section*{Biographies}

\textsc{Bin Li} (bin.li@nuist.edu.cn) is an Associate Professor at Nanjing University of Information Science and Technology. He received the Ph.D. degree from Beijing Institute of Technology in 2019. From 2013 to 2014, he was a Research Assistant at The Hong Kong Polytechnic University, Hong Kong.
From 2017 to 2018, he was a Visiting Student at University of Oslo, Norway.
His research interests include unmanned-aerial-vehicle communications, intelligent reflection surface and mobile edge computing.

\textsc{Wancheng Xie} (xiewancheng@stu.xmu.edu.cn) 
is currently pursuing the M.S. degree with the Institute of Artificial Intelligence, Xiamen University, Xiamen, China. His current research interests include joint sensing, communication and computation, deep reinforcement learning, and mobile edge computing.

\textsc{Zesong Fei} (corresponding author, feizesong@bit.edu.cn) 
is a Professor at Beijing Institute of Technology (BIT). He received the
Ph.D. degree in electronic engineering from BIT in 2004. He is also the
Chief Investigator of the National Natural Science Foundation of China. He has authored or coauthored over 60 articles in IEEE journals. His research
interests include wireless communications and multimedia signal processing.

\end{document}